# Upconversion Dual-comb Passive Spectroscopy of Mid-IR Incoherent Light

**Mingchen Liu[1,*], Matthew Heyrich[1,2], Sichong Ma[1,2], and Scott A. Diddams[1,2]**

*[1] Department of Electrical, Computer, and Energy Engineering, University of Colorado Boulder, Boulder, CO 80309, USA.*
*[2] Department of Physics, University of Colorado Boulder, Boulder, CO 80309, USA.*
**mingchen.liu@colorado.edu*

**Abstract:** We introduce cross-comb correlation spectroscopy, which correlates signals from asynchronous sum-frequency sampling of mid-IR incoherent light with two near-IR combs. We experimentally measure the frequency of phase-randomized mid-IR light with 100-MHz resolution using near-IR detectors. 

Dual-comb spectroscopy (DCS) has become a powerful alternative to conventional Fourier-transform and dispersive spectrometers, with better speed, resolution, and accuracy. Standard DCS interrogates samples actively with coherent combs. However, most naturally occurring optical sources are incoherent and many objects cannot always be accessed by active illumination. Dual-comb correlation spectroscopy (DCCS) was developed to extend comb-based spectroscopy to passive sensing with incoherent light while preserving key advantages of DCS [1–3].

A remaining limitation is that DCCS measures radiation only within the spectral region covered by the dual combs. This constraint is particularly restrictive in the mid-infrared (MIR), where important molecular and thermal signatures reside. Although MIR comb technologies have advanced considerably, generating two mutually coherent combs in longer-wavelength regions remains much more challenging than in the near-infrared (NIR). In addition, photodetectors at longer infrared wavelengths generally exhibit lower sensitivity, higher noise, and more demanding operating requirements than their NIR counterparts, imposing a practical bottleneck on such measurements.

To address these limitations and enable dual-comb passive spectroscopy of incoherent light at longer wavelengths, here we introduce Cross-comb Correlation Spectroscopy (CCCS), as illustrated in Fig. 1a. The MIR incoherent light (target) is first split into two branches and upconverted to NIR via sum frequency generation (SFG) with two NIR combs, referred to here as upconversion combs, of slightly different repetition rates. In each branch, the upconverted light is then combined with a heterodyne comb and detected on a NIR balanced photodetector (BPD). The two BPDs outputs are digitized and mixed to obtain a correlation interferogram. The interferograms can be averaged and then Fourier transformed to recover the spectral information of the MIR incoherent light.

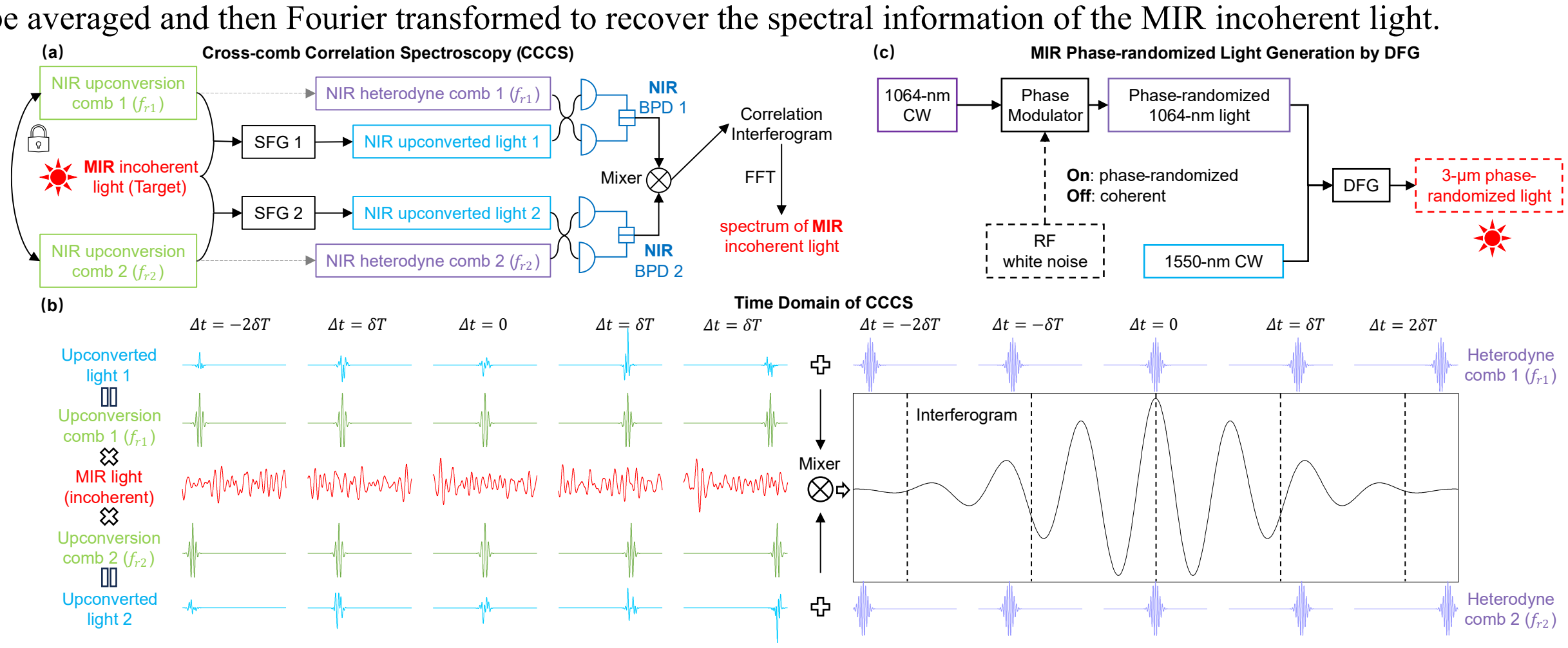

Fig.1 (a) Schematic of CCCS. (b) Time domain picture of CCCS. (c) Artificial phase-randomized MIR light generation.

Figure 1b illustrates the time-domain picture of CCCS. The MIR incoherent light is first nonlinearly sampled by two NIR pulse trains (upconversion combs) at different delays, and the upconverted fields are subsequently linearly heterodyned with two NIR pulse trains (heterodyne combs). In each branch, the heterodyne comb shares the same comb parameters ($f_{rep}$ and $f_{ceo}$) and timing as the corresponding upconversion comb. This is because the heterodyne comb is generated as a different spectral part of the same comb as the upconversion comb, similar to standard coherent cross-comb spectroscopy (CCS) [4-5]. The slowly varying delay between the two sampling pulse trains, $\Delta t$, set by their repetition-rate difference, provides an effective time-delay scan analogous to that in FTIR. This produces an autocorrelation signal of the MIR incoherent light, modified by the temporal envelope of the NIR comb pulses. Overall, the time-domain picture resembles that of DCCS, but with an additional upconversion step that helps and modifies, rather than prevents, spectral recovery of the MIR incoherent target.

To experimentally demonstrate the concept of CCCS, we generate artificially phase-randomized MIR light around 3 µm as the detection target through difference-frequency generation (DFG) between a 1550-nm continuous-wave (CW) laser and a 1064-nm CW laser. As shown in Fig. 1c, incoherence is emulated by sending the 1064-nm laser through a phase modulator driven by RF white noise, which randomizes its optical phase. This phase randomness is transferred to the generated 3-µm field. Two mutually coherent 1560-nm, 100-MHz Er:fiber frequency combs serve as the upconversion combs, with their repetition-rate difference set to 400 Hz. Each heterodyne comb is obtained by bandpass filtering the supercontinuum generated in a highly nonlinear fiber pumped by the corresponding upconversion comb. The two BPDs are standard NIR InGaAs detectors.

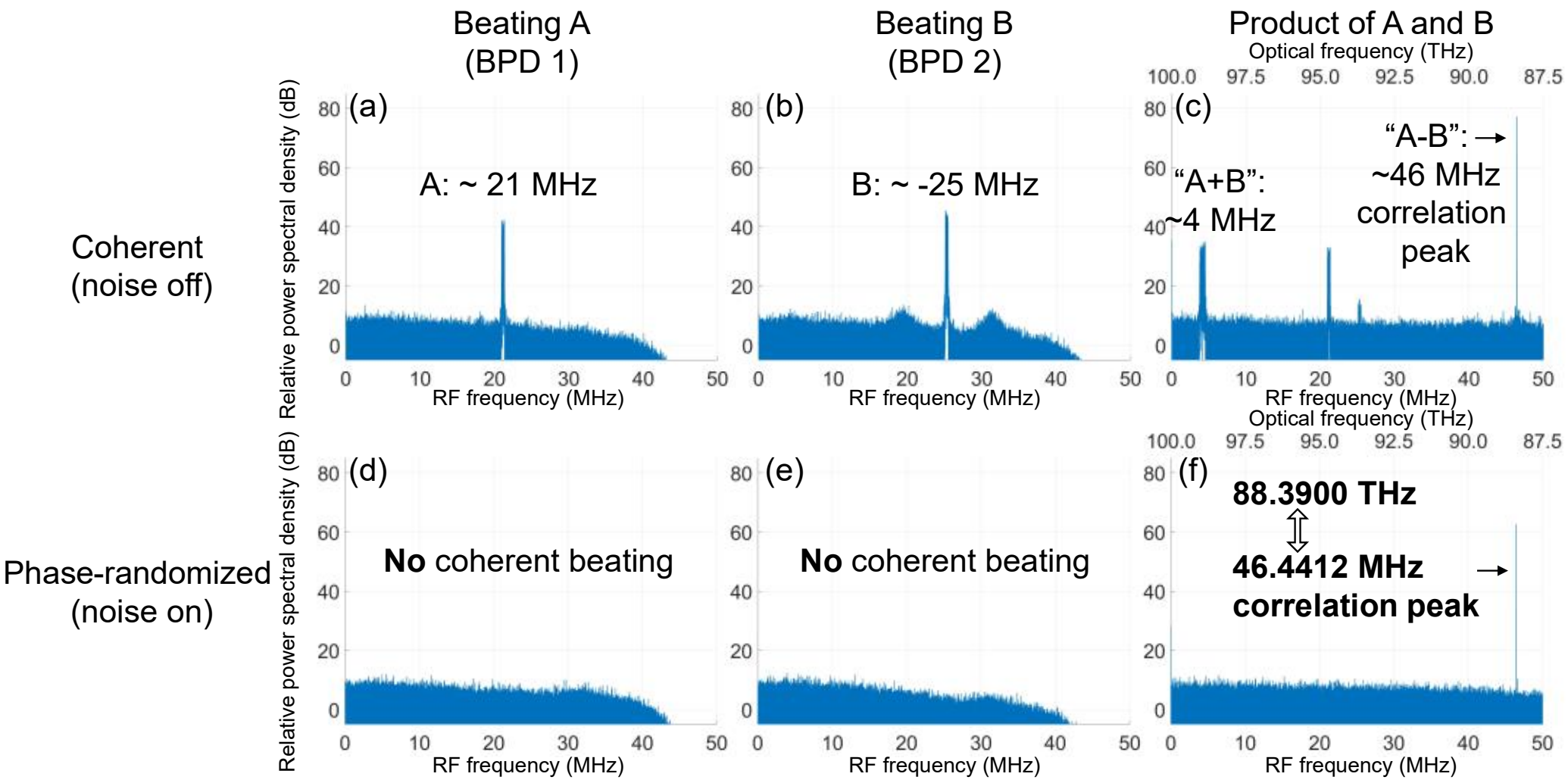

Fig.2 Frequency-domain comparison of the two heterodyne signals and their product for coherent and phase-randomized MIR detection targets.

Figure 2 shows the frequency-domain results for the coherent case (Fig. 2a–c, white-noise drive off) and the phase-randomized case (Fig. 2d–f, white-noise drive on). In the coherent case, each BPD records a strong heterodyne peak with a finite linewidth, corresponding to the coherent 3-µm CW field. Multiplying the two BPD outputs produces additional sum- and difference-frequency components, apart from residual leakage of the individual heterodyne signals. The difference-frequency component is the correlation peak that encodes the MIR target frequency. In this limit, the system operates as a nonlinear analogue of a dual-comb Vernier spectrometer [6], enabling measurement of a coherent MIR CW field over a large unambiguous frequency range with a high resolution limited by the NIR comb linewidths. While this coherent measurement is interesting as a separate direction and useful for system validation, it does not by itself establish CCCS for incoherent-light detection.

In the phase-randomized case, no discrete peak is observed in either individual BPD spectrum, although the detector voltages are clearly visible in the time domain. This is because the randomized target phase is transferred to the heterodyne signals, making them noise-like in the frequency domain and spread over a broad bandwidth of the applied white-noise modulation. Despite the absence of visible single-detector peaks, the two noise-like signals remain correlated, producing a strong peak in the spectrum of their product. Using the known $f_{rep}$ and $f_{ceo}$ of the two upconversion combs, the RF correlation peak at 46.4412 MHz is mapped back to an MIR optical frequency of 88.3900 THz with a frequency resolution of 100 MHz, set by $f_{rep}$. This result agrees well with a commercial MIR FTIR measurement of the target, confirming the feasibility of CCCS for phase-randomized MIR light.

In summary, we propose and experimentally demonstrate CCCS by measuring phase-randomized MIR light with NIR photodetectors, circumventing the challenges of mutually coherent MIR comb generation and direct photodetection. Ongoing work aims to extend CCCS to real incoherent sources, including amplified spontaneous emission (ASE) and sunlight, and thermal emission from natural objects.

**Acknowledgements**
This work was supported by the W. M. Keck Foundation and the AFOSR. MH acknowledges support from the NSF Graduate Research Fellowship Program.